# EXTENDED FAST SEARCH CLUSTERING ALGORITHM: WIDELY DENSITY CLUSTERS, NO DENSITY PEAKS

Zhang WenKai[1] and Li Jing[2]

[1]School of Computer Science and Technology, University of Science and Technology of China, Hefei, 230026, China
zwk@mail.ustc.edu.cn
[2] School of Computer Science and Technology, University of Science and Technology of China, Hefei, 230026, China
lj@ustc.edu.cn

## ABSTRACT

*CFSFDP (clustering by fast search and find of density peaks) is recently developed density-based clustering algorithm. Compared to DBSCAN, it needs less parameters and is computationally cheap for its non-iteration. Alex. at al have demonstrated its power by many applications. However, CFSFDP performs not well when there are more than one density peak for one cluster, what we name as "no density peaks". In this paper, inspired by the idea of a hierarchical clustering algorithm CHAMELEON, we propose an extension of CFSFDP, E_CFSFDP, to adapt more applications. In particular, we take use of original CFSFDP to generating initial clusters first, then merge the sub clusters in the second phase. We have conducted the algorithm to several data sets, of which, there are "no density peaks". Experiment results show that our approach outperforms the original one due to it breaks through the strict claim of data sets.*

## 1. INTRODUCTION

Clustering is known as the unsupervised classification in pattern recognition, or nonparametric density estimation in statistics [1]. The aim is to partition given data set of points or objects into natural grouping(s) according to their similarity to improve understanding on the condition of no priori-knowledge, or be as a method to compress data. Cluster analysis has been widely used in a lot of fields, like computer version ([2], [3], [4]), bioinformatics ([5], [6], [7]), image progressing ([8], [9], [10], [11]), Knowledge Discovery in Databases (KDD), and many other areas ([12]). Thousands of clustering algorithms have been proposed, challenges still remain: differing shapes, high dimensions, how to determine the clusters number, how to define a right clustering, hard to evaluate.

Density-based clustering algorithms which classify points by identifying regions heavily populated with data, such as DBSCAN [13] and GDBSCAN [14], OPTICS [15], and DBCLASD [16], have performed well while handling problems of arbitrary shapes of subclasses. DBSCAN [13] is a representative of density-based methods, by the definition of core points, density connection, in which clusters defined as high density regions separated by low density regions in the feature space can be detected without the need for clusters number. However the appropriate threshold MinPts(minimum number of points) for distinguishing core points from border points is hard to select, with a high MinPts, thin clusters(relative low density) would be ignored.

Similar to DBSCAN [13], recently, CFSFDP (clustering by fast search and find of density peaks) [17] was proposed by Alex and Anlessandro to detect non-spherical groups, which does not need to pre-specify the number of clusters of variant shapes either. In addition, CFSFDP needs less parameters. CFSFDP finds the clusters of points by a two phase progression. During the first phase, CFSFDP uses a well-designed decision group to find out the cluster centers, so-called density peaks. During the second phase, each remaining point is assigned to the same cluster as its nearest neighbor of higher density. In a DBSCANs perspective, CFSFDP assumes that every object is density-connected with its nearest neighbor of higher density. Compared to mean-shift methods such as [18], [19], CFSFDP [17] is computationally cheaper for the procedure of maximizing the density field for each data point in the mean-shift approach. By the experiments of identifying the number of subjects in the Olivetti Face Database [20], the team have shown CFSFDP's capacity to solve high dimensional data [17].

However, in our opinion, there are some drawbacks of the beautiful CFSFDP [17], which will limit the application of CFSFDP. Firstly, just as DBSCAN [13], thin clusters would not be captured by the decision graph. Besides, a rigid hidden requirement for getting right clusters is that, each cluster in the data sets must have a density peak and only one peak is promised, otherwise CFSFDP will split natural groups. In this paper, inspired by CHAMELEON [21], we present a novel hierarchical clustering algorithm based on CFSFDP. Thus our approach can find thin clusters. In addition, it eliminates the strict claim of density peaks. To display our efforts, we benchmark our algorithm on the data sets draw from [21], [22], [23], of which there is no unique density peak for each cluster. Our technique gets partitions of these data sets as well as that generated by the methods proposed in the papers where the data set was designed.

We discuss details of CFSFDP [17], CHAMELEON [21] in Section 2. In Section 3, we present drawbacks of CFSFDP and our efforts to overcome these limitations. Section 4 describes our algorithm in detail. In Section 5, we benchmark the performance of our approach on some data sets from other literatures. Finally, a conclusion and direction of future works are shown in Section 6.

## 2. BASIC CONCEPTS

This Section presents two methods and some concepts involved in our technique, what are necessary to understand our approach. If one has been familiar to CFSFDP and CHAMELEON, he could skip to Section 3, where we discuss some disadvantages of CFSFDP in details.

### 2.1. CFSFDP

CFSFDP [17] generates clusters by assigning data points to the same cluster as its nearest neighbour with higher density after cluster centres are selected by users. The cluster centres are defined as local density maxima, Alex and Alessandro designed a heuristic method for customers to select the genuine cluster centres, what is named as decision graph.

Two important quantities are considered in the decision group: local density $\rho_i$ of each point $i$, its distance $\delta_i$ from points of higher density. Definition of $\rho_i$, $\delta_i$ follows:

**Definition 1**: The density of a point $i$, denoted by $\rho_i$, is defined as
$$\rho_i = \sum_j \chi(d_{ij} - d_c) .  \tag{1}$$
Where $\chi(x) = 1$ if $x < 0$ and $\chi(x) = 0$ otherwise, and $d_c$ is a cutoff distance, which is the only parameter need to be determined by customers. In principle, $\rho_i$ equals to the number of points which are closer than $d_c$ to point $i$.

**Definition 1**: The minimum distance of point $i$ from any other point of higher density, denoted by $\delta_i$, is computed by
$$\delta_i = \min_{j:\rho_j > \rho_i} d_{ij} .  \tag{2}$$

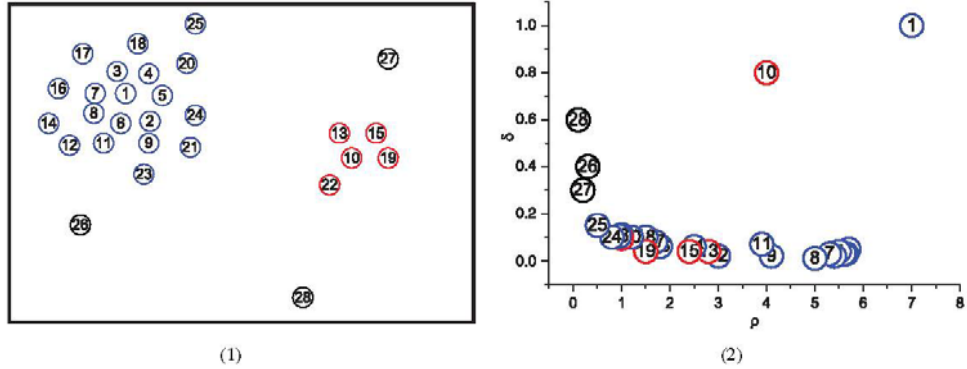

Figure 1. The CFSFDP in two dimensions. (1) Point distribution. Data points are ranked in order of decreasing density. (2) Decision graph for the data in (1). Different colors correspond to different clusters. [17]

Notice that, for the point of largest density, $\delta_i$ is redefined as $\delta_i = \max\limits_{j \in dataset} d_{ij}$. The simple observation that points of high local density and high density distance are local density maxima, namely density peaks or cluster centres, is the core of this procedure to select cluster centers.

To identify density peaks defined as below, a method named as "decision graph" is introduced to help users to make a decision. Basically, decision graph is a figure plotting $\delta_i$ as a function of $\rho_i$, is illustrated by the two-dimensional simple example in Fig.1. Points 1 and 10 are the only two points of high $\delta$ and high $\rho$, as a result, they are the cluster centers. Points 26, 27, 28 can be considered as outliers for a relatively high $\delta$ and low $\rho$ (which indicates that they are isolated points).

After cluster centers have been found, CFSFDP allocates the rest points to the same cluster as its nearest neighbourhood with higher density.

## 2.2. CHAMELEON

CHAMELEON discovers clusters of a given data sets gradually by finding groups consisted of most similar points. There are 3 main steps in CHAMELEON: create the k-nearest Neighbour Graph according to data points, partition the graph into sub-classes, then merge the subsets. It's common to model data items as a graph in agglomerative hierarchical clustering techniques [1], CHAMELEON models data based on the widely used k-nearest neighbour graph technique. After the graph was built, a efficient graph partitioning algorithm mMETIS [24] is used to find the initial sub-clusters. As criteria to aggregate the sub clusters, CHAMELEON computes the relative interconnectivity $\mathrm{RI}(C_i, C_j)$ and relative closeness $\mathrm{RC}(C_i, C_j)$ between each pair of clusters $C_i$ and $C_j$. In the merge phase, each round the cluster pair of highest

$$\mathrm{RI}(C_i, C_j) \times RC(C_i, C_j)^\beta \qquad (3)$$

will be merged, where $\beta$ is a user defined parameter to give different importance to the two criterions. The merge progress will be stop when cluster number is equal to the number predefined or there is no cluster pair of which the value of (3) is bigger than user determined threshold.

In (3), the relative inter-connectivity $\mathrm{RI}(C_i, C_j)$ between cluster $C_i$ and cluster $C_j$ is defined as the sum of the weight of edges span the two clusters normalized with respect to the internal inter-connectivity of cluster $C_i$ and cluster $C_j$. The internal inter-connectivity of cluster is computed by adding the weight of edges that partition the cluster into two roughly equal parts. Meanwhile CHAMELEON defines the relative closeness $\mathrm{RC}(C_i, C_j)$ of cluster $C_i$ and cluster $C_j$ as the average weight of edges span the two clusters, which also is normalized with respect to the internal closeness of each cluster.

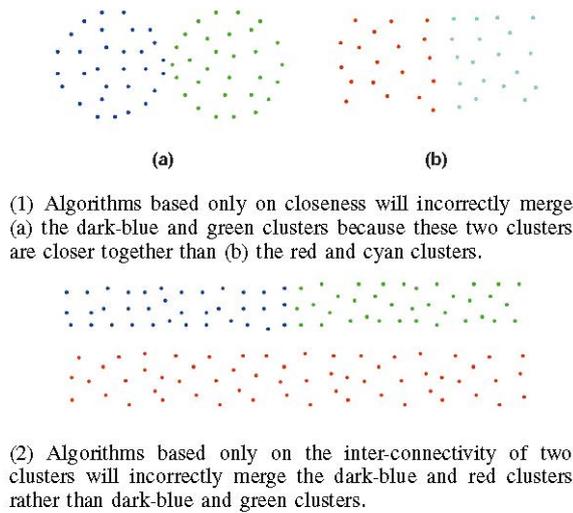

(1) Algorithms based only on closeness will incorrectly merge (a) the dark-blue and green clusters because these two clusters are closer together than (b) the red and cyan clusters.

(2) Algorithms based only on the inter-connectivity of two clusters will incorrectly merge the dark-blue and red clusters rather than dark-blue and green clusters.

Figure 2. Clustering considering either closeness or inter-connectivity. [21]

The key difference of CHAMELEON and other hierarchical clustering algorithms like CURE [25], ROCK [26] is that it accounts for both inter-connectivity and closeness while identifying the most similar pair of clusters in the third step [21]. The disadvantages of only considering either closeness or inter-connectivity is these schemes could merge wrong pair of clusters. A sample of that was given by George, as shown in Fig.2. In Fig.2.(1), algorithms only based on closeness (CURE and related schemes) merge incorrectly because through clusters of (a) — the dark-blue and green clusters — are closer to each other than the case of (b) — the red and cyan clusters, clusters of (b) connect better than those of (a). In Fig.2.(2), the inter-connectivity of the dark-blue cluster and the red cluster is bigger than that of the dark-blue cluster and the green cluster, however the dark-blue cluster is closer to the red cluster than the green one. Thus an algorithm based only on the interconnectivity (ROCK [26]) merge the dark-blue cluster with the red cluster incorrectly.

## 3. LIMITATIONS OF CFSFDP AND OUR CONTRIBUTION

In this Section, firstly we discuss our understanding towards limitations of CFSFDP in theory. To help illustrate the limitation intuitively, an instance is presented.

CFSFDP has got high performance classifying several data sets, the success of these examples verifies the availability of CFSFDP in some degree. As shown in Section 2, it generates groups by identifying clusters with density maxima. However, it is impossible to get natural clusters by CFSFDP when the local densities of data points in some or all natural clusters of the data sets is random distributed, such that instead of one density peak two or more density peaks appear in one cluster. In such a case, it is hard for CFSFDP to pick up all the reasonable cluster centers. What's more, even if a reasonable cluster center set is found by CFSFDP, the natural cluster would be split. Reasonable cluster center here means that a point as density maximal, with which, the result cluster of CFSFDP clustering procedure (assign each point to the same cluster of its nearest neighbour of higher density than itself) is not consisted of parts of different natural clusters. Reasonable cluster center set means a set of reasonable cluster centers. The cluster center set presented in Fig.4.B, is an unreasonable cluster center set for the red cluster is composed of part of the "arc" cluster and part of left cluster.

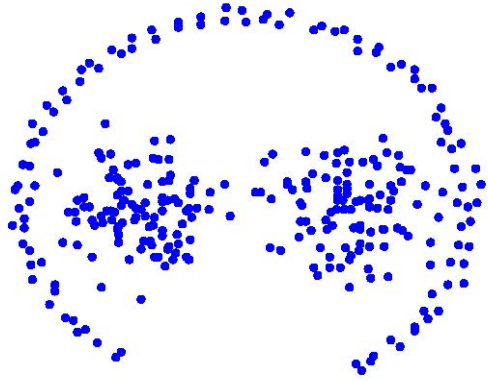

Figure 3. The data set drawn from [22]. There are 3 native clusters of diverse density, complex shapes, two of the three are surrounded by the third one.

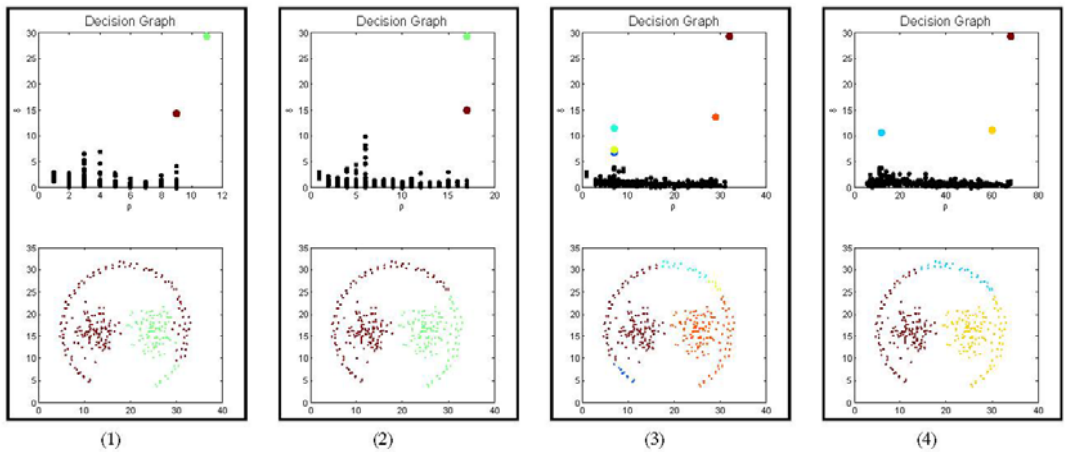

Figure 4. CFSFDP groups the data set in Fig.3, Different colors correspond to different clusters, the figure above in each sub figure is the decision graph, the figure below is the correspond dividing result. (1)$d_c = 1$:101 such that the largest density equals to about 1% of the number of points size, 2 clusters. (2) $d_c = 1$:540 such that the largest density is equal to about 2% of the size of data set, 2 clusters. (3) $d_c = 2$:250, such that the largest density equals to about 4% of the number of total points, 5 clusters. (4) $d_c = 3$:814, such that the largest density equals to about 10% of the number of total objects, 3 clusters.

Let's consider the test case in Fig.3. The data set is drawn from [22], there are 3 natural clusters in the data set: the left cluster surrounded by the arc, the right cluster surrounded by the arc, and the arc cluster. We test CFSFDP on this data set by variant dc, the clustering result of some values (1.101, 1.540, 2.250, 3.814) is shown in Fig.4. Except the experiments presented in Fig.4, many other values of dc have been tested, unfortunately CFSFDP can't divide the points into natural groups with any value of $d_c$. In particular, CFSFDP makes mistake on the arc cluster, parts of which are resigned into wrong clusters.

Firstly, poor ability of decision graph to select cluster centers is one reason of such a bad outcome, because performance of CFSFDP is highly sensitive to the procedure determining cluster centers. [17] says, cluster centers of relative high local density $\rho$ & relative high $\delta$ are of remarkable location in the decision graph, and indeed its true for some cases, as shown in Fig.1 where cluster centers are of high $\rho$ and high $\delta$ in global. Nevertheless, when clusters of more complex data sets, either thin cluster centers of relative low $\rho$ and high $\delta$, or other cluster centers of relative high $\rho$ and low $\delta$, would be overlooked easily. As an example, in Fig.4, the relative thin cluster centers belong to the "arc" cluster are easily ignored for we only select the points of remarkable location as cluster centers. One may argue that with a more suitable value of $d_c$, density peaks would be prominent in decision graph and be easy to point out by users. In our opinion, it may be or may not be, it depends on the data sets. A good method to select cluster centers should reduce its dependency upon data sets as much as it can, such that it would detect diverse density or diverse $\delta$ cluster centers.

Another cause of wrong clusters detected by CFSFDP is that CFSFDP divides points based on cluster centers, such that CFSFDP might divide the natural cluster if there are more than one cluster center was determined in a natural cluster. For example, Fig.4 shows the clustering consequence of the [22] data set after choosing reasonable cluster center set by our new decision graph. In Fig.4, points of the arc cluster which is split into several clusters. Inspired by the clustering progress of hierarchical clustering algorithm, a novel clustering algorithm consisted of two phases brings out to us: 1) generate sub classes by CFSFDP, 2) merge the sub classes by the similarity between clusters.

### 3.1. Modelling the Similarity between Clusters

To break through the barriers of agglomerative hierarchical approaches discussed in Section 2, our algorithm looks at their Relative Inter-connectivity $\mathrm{RI}(C_i, C_j)$ and their Relative closeness $\mathrm{RC}(C_i, C_j)$ while merging cluster $C_i$ and cluster $C_j$, which is similar with the aggregation phase of CHAMELEON. By considering both of these criteria, our scheme selects clusters that are well connected as well as close together to merge. However, there is a shortage in CHAMELEON that is CHAMELEON, which models sub clusters based on the widely used k-nearest neighbor graph, would fail to merge the correct cluster pair in some cases. To solve that, we using a variant of k-nearest neighbor graph to model the sub clusters. Other than the difference of the model to represent sub clusters, we compute the relative inter-connectivity and the closeness almost the same as CHAMELEON, we also use the value defined by (3) to model the similarity between clusters.

In this Section, firstly we show the detail of the relative inter-connectivity and the relative closeness. In the remainder, the detail of drawback in CHAMELEON's merging phase is presented, then our solution is followed.

The relative inter-connectivity $\mathrm{RI}(C_i, C_j)$ between cluster $C_i$ and cluster $C_j$ is given by

$$\mathrm{RI}\left(C_i, C_j\right) = \frac{EC_{\{c_i, c_j\}}}{\frac{|c_i|}{|c_i| + |c_j|} EC_{c_i} + \frac{|c_j|}{|c_i| + |c_j|} EC_{c_j}} . \tag{4}$$

Where $EC_{\{C_i, C_j\}} = \sum E(u, v)$ is sum of weight of edges connecting the two clusters. $\mathrm{E}(u, v)$ is defined as (4), this is the main difference between CHAMELEONs merging phase and our merging phase.

The Relative closeness $\mathrm{RC}\left(C_i, C_j\right)$ is computed by

$$\mathrm{RC}\left(C_i, C_j\right) = \frac{\bar{S}_{EC_{\{C_i, C_j\}}}}{\frac{|c_i|}{|c_i| + |c_j|} \bar{S}_{EC_{c_i}} + \frac{|c_j|}{|c_i| + |c_j|} \bar{S}_{EC_{c_j}}}. \tag{5}$$

Where $\bar{S}_{EC_{\{C_i, C_j\}}} = \frac{EC_{\{C_i, C_j\}}}{|E(u, v)|}$ is the average weight of edges span the two clusters. To get the value of inner criteria, as $EC_{C_i}$ or $\bar{S}_{EC_{C_i}}$, CHAMELEON takes graph partitioning technology to bipartite the target cluster, while we split the target cluster by CFSFDP. In particular, at first, we utilize the CFSFDP clustering algorithm with the novel decision graph(described in previous part of this

Section) to clustering the points of current cluster into two sub classes; then we compute F based on the clustering result. There are two advantages of taking use of CFSFDP rather than graph partitioning methods to split the target cluster. Firstly, CFSFDP is very fast with specific cluster centers, the time complexity is ○ (n), $k$ is the number of points of the target cluster. The selection of cluster centers can be done in constant time based on the computation in the initial clustering phase. Secondly, we use CFSFDP in all phases of our clustering method to keep consistency in theory, in our view, that meets the famous Ockham's Razor principle.

CHAMELEON models the similarity based on the widely used $k$-nearest neighbor graph. The benefit of representing data as a $k$-nearest neighbor graph covers that simplifying topology of data sets by disconnecting the points are far apart, capturing the concept of neighbourhood dynamically, and there are many methods offered to handle graph data [21]. However, in the merge phase, we find that CHAMELEON based on the standard k-nearest graph would fail to merge the correct cluster pair in some cases. The example in Figure.5.(1) illustrates, an algorithm that considering the inter-connectivity as CHAMELEON (presented in Section 2) will prefer to merge incorrectly the red cluster with the blue cluster, rather than with the green cluster. Notice that, because the blue cluster is much sparser than the red cluster, as a result, the measure of connectivity and closeness depends on point $P$, and almost all the $k$-nearest neighbours of point $P$ belong to the red cluster, leading to that both the criteria (inter-connectivity and closeness) of the red cluster and the blue cluster are higher than those of the red cluster and the green cluster.

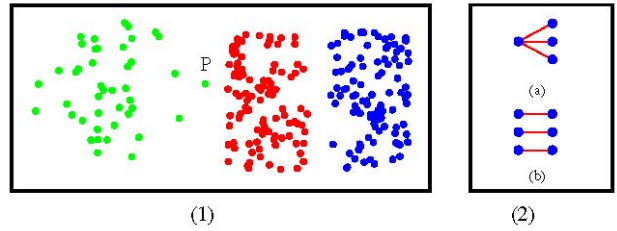

Figure 5. (1) Chameleon prefers to merge incorrectly the red cluster with the blue cluster rather with the green cluster. (2) A simplified instance to explain why Chameleon fails in (1).

Further, we find that CHAMELEON fails because it is almost impossible for CHAMELEON based on the standard k-nearest graph to distinguish case (a) from case (b) in Figure.5. (2). Where the weight of edges is nearly the same. In figure.5, we ignoring the normalization because the inner criteria are the same in this example.

To address that disadvantage, we bring out a variant of $k$-nearest neighbour graph to model the sub clusters. In this paper, the edge connecting point of cluster $C_i$ with point of cluster $C_j$ is given by

$$\mathrm{E}(u,v) = \max\left(\left\{E(u,p) \middle| u \in C_i \cap p \in C_j \cap u \in KN(p) \cap p \in KN(u)\right\}\right). \quad (6)$$

Where $\mathrm{E}(u,v)$ represents the edge connecting point $u$ with point $v$. $KN(\cdot)$ is an operator to get the k-nearest neighbor. The $\max(\cdot)$ function is to get the edge of maximum weight, the weight here means the similarity. By only considering the edge of maximum weight, our approach can distinguish the two type k-nearest neighbor relationship. In figure.5. (2). (b), we remove the two slash edges, such that, we can get a smaller inter-connectivity of case (a) than that of case (b). Even if there is still a similar closeness, it is possible to distinguish case (a) from case (b) by giving a higher importance to the inter-connectivity, by specifying a value smaller than $1.o$ to $\beta$ in (3).

# 4. THE EXTENDED CFSFDP (E_CFSFDP)

| **Algorithm 1 E_CFSFDP** $(X, d_c, N_{neighbor}, \beta)$ |
|---|

**Inputs:**

$X = \{X_1, X_2, X_3, \ldots, X_n\}$ Set of data objects

$d_c$ : radius to compute ρ.

$N_{neighbor}$: value of neighbour number while modelling the cluster similarity.

β : parameter to control the importance of the two criteria in (3).

**Output:**

$C = \{C_1, C_2, C_3, \ldots, C_n\}$ Set of clusters

{**Phase I**} find initial sub-clusters by CFSFDP

1: compute the similarity matrix.

2: compute the local density $\rho$ and the distance $\delta$ of each object.

3: select cluster centers by decision graph.

4: allocate each point to the same cluster as its nearest neighbour of higher $\rho$.

{**Phase II**} merge the sub-clusters

5: construct the model of sub-clusters.

6: compute the $\mathbf{RI(C_i, C_j)} \times \mathbf{RC(C_i, C_j)}^{\beta}$ matrix for each sub-clusters pair $\{C_i, C_j\}$ based on the model built in step 4.

7: merge the cluster pair of highest $\mathbf{RI(C_i, C_j)} \times \mathbf{RC(C_i, C_j)}^{\beta}$.

8: repeat step 6 and step 7 until the termination conditions are meet.

## 4.1. The Description of E_CFSFDP

While CFSFDP algorithm needs one parameter $d_c$, our algorithm E_CFSFDP requires three parameters for the extensions presented in Section 3. As in CFSFDP, $d_c$ is a value to compute the local density ρ in (1). [17] gives an empirical hint to specify $d_c$, choosing the value so that the average number of neighbours is around *1%* to *2%* of the total number of points in the data set. $N_{neighbor}$ is the number of neighbours for modelling the similarity of sub-clusters. β is the factor to control the relative importance of inter-connectivity and closeness in the merge step.

We describe the steps of our algorithm E_CFSFDP in Algorithm 1.

Our algorithm starts at getting the similarity matrix of the data set. Many methods have been proposed by the machine learning community to measure the similarity between different objects, such as the common used Euclidean Distance, Mahalanobis Distance, Cosine Distance, SNN Similarity. One can choose appropriate method based on the application, i.e., Euclidean Distance for 2D data set, SNN Similarity for high dimensional data set. After that, we initialize the clustering group by CFSFDP. What's different from the standard CFSFDP, our scheme chooses as more cluster centers as possible in step 3 to overcome the limitation of CFSFDP described in Section 3.

In Phase II, the algorithm computes the value of necessary variables for merging in Step 5 and Step 6. In Step 5, the variant k-nearest neighbour graph is constructed, where algorithms based on k-d trees can be used [27]. In Step 7, E_CFSFDP combines two clusters into one in each iteration. To do so, the $RI(C_i, C_j) \times RC(C_i, C_j)^\beta$ matrix needs to update in each loop, and the renewal of value refer to the sub cluster has been merged is necessary.

As a hierarchical algorithm, E_CFSFDP will not terminate until there is only one cluster remained. However, in real application, that seems meaningless. So in real application, E_CFSFDP usually terminates at the moment when the number of clusters is equal to the user specified parameter k. And CHAMELEON [21] proposes an alternative scheme to terminate the combination, that is to set two user input threshold $T_{RI}$, $T_{RC}$, thus the algorithm stops when there is no cluster pair, of which the inter-connectivity $RI$ and the closeness $RC$ are bigger than the respond threshold $T_{RI}$ and $T_{RC}$. While in the experiments of this paper, we stop the algorithm by a value $k$ of genuine clusters' number.

## 4.2. Performance Analysis

The runtime complexity of E CFSFDP mainly depends on the two phases in the scheme. In the initial phase, E_CFSFDP is almost the same as CFSFDP.

Let´s consider the runtime of CFSFDP. CFSFDP needs $\circ(n^2)$ to compute local density and the distance where $n$ is the number of objects in the input data set. For the progress to determine the cluster centers, we take no account of the time for the user to select cluster centers due to that is hard to quantize correctly. CFSFDP needs $\circ(n)$ to construct the decision graph. After the cluster centers are chose, CFSFDP assigns each point to the same cluster as by the order of descend local density, thus CFSFDP needs $\circ(n\log n)$ to sort the points with quick sort [28], and CFSFDP's assignment procedure costs $\circ(n)$, such that, the total time complexity of CFSFDP is $\circ(n^2)$ if the similarity matrix has been finished. In the first phase, E_CFSFDP costs the same time as the overall time of original CFSFDP, indicates the time complexity of E_CFSFDP's first phase is $\circ(n^2)$. Moreover, our analysis will focus on the computation of inter-connectivity and relative closeness which have been described in Section 3, for the main cost of E_CFSFDP's second phase comes from the computation and the update of these criteria.

The computation of the two criteria is based on the variant -nearest neighbor graph, which can be constructed based on the basic $k$-nearest neighbor graph. For low dimensional data set, the amount of time to construct the basic $k$-nearest neighbor graph is $\circ(n\log n)$ if algorithms based on k-d trees [27] are used. For high dimensional data set, schemes based on $k$-$d$ trees are not applicable ([29], [30]), leading to an overall time complexity of $\circ(n^2)$. To get the variant model presented in Section 3, our algorithm need to check all $k$ nearest neighbours for each item, leading to the runtime complexity is $\circ(kn)$, $k$ is the number of neighbours.

To simplify the analysis, we assume that each cluster is of the same size, thus if the number of sub clusters is $m$, the size of each cluster is $m/n$. E_CFSFDP needs to traverse part of the variant $k$-nearest neighbor graph to compute the interconnectivity and relative closeness for each cluster pair, of which data points belong to the cluster pair, resulting in time complexity of $\circ(2n/m)$. The amount of time of normalizing the inter-connectivity or the closeness is also $\circ(2n/m)$ for the bisection costs the same time as that of CFSFDP's alignment progress. In Step 5, there are $\binom{n}{2}$ cluster pairs to be solved, causing the amount of time $\circ(0.5n^2 \times 2n/m) = \circ(mn)$. While merging iteratively, the runtime depends on the time of finding the most similar cluster pair and updating the criteria. By using a heap-based priority queue, the amount of time required to find the most similar cluster pair is $\circ(m^2\log m)$. The time to update the matrix is $\circ(n)$ for there are most $m-1$ steps, each step requires the time of $\circ(2n/m)$.

Thus, the overall complexity of the extended CFSFDP (E_CFSFDP) is $\circ(n^2 + n\log n + mn)$. Compared to the runtime of original CFSFDP $\circ(n^2)$, the largest term $n^2$ of the two approaches both comes from the computation of local density, the amount of time of our algorithm doesn't increase much, especially for a large value of $n$.

## 5. EXPERIMENTS

In order to demonstrate the performance of our algorithm E_CFSFDP, we benchmark it on several 2D data sets consist of clusters of diverse density, shapes, which have not no unique density peak for each cluster. The reason why we adopt 2D data sets to benchmark our scheme is that it is easy to visualize the clustering results of 2D data sets, making the comparison of different algorithms much easier. Our experimental study focuses on the comparison of the basic CFSFDP and its extension E_CFSFDP proposed by us, to testify that E_CFSFDP can handle the data set, in which there is no unique density peak for each cluster.

This paper measures the similarity between data points with the famous Euclidean Distance, which is used widely to measure the similarity of spatial data, 2D or 3D. And though we determine parameters ($d_c$, $N_{neighbor}$, $\beta$) based on the application, we will show that our algorithm E_CFSFDP is robust with respect to parameters by presenting the clustering results of E_CFSFDP with different combinations of value of $d_c$, $N_{neighbor}$, and β. Besides, for convenience, we use percentage to denote the value of parameter $d_c$, which is used in (1) to compute the local density, i.e., we say $d_c = 2\%$, indicates that with the $d_c$, s. t., $\max \rho = 2\% \times size(dataset)$.

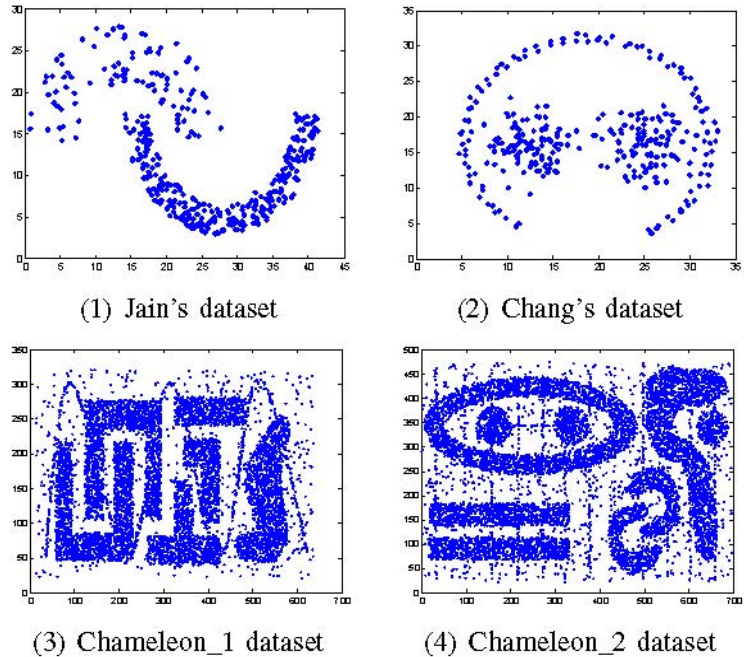

(1) Jain's dataset        (2) Chang's dataset

(3) Chameleon_1 dataset        (4) Chameleon_2 dataset

Figure 6. 2D data sets to benchmark E_CFSFDP

### 5.1. Data Sets

Four data sets proposed by other papers are used to evaluate our algorithm in this paper. In particular, Jain's dataset is taken from [23], Chang's dataset is taken from [22], Chameleon_1 and Chameleon_2 are drawn from [21]. As presented in Fig.6, clusters of the 4 data sets are of diverse density, complex shapes. The most important feature of these data sets shared is no obvious single density peak for most clusters, even in Chang's data set, there is one cluster — the arc one — which don't have a unique density peak. We will show you that our algorithm performs better to deal with this case below.

### 5.2. The Evaluation of Clustering Results

### 2.6.1. Jain's data set

As shown in Fig. 6, there are 2 clusters, 373 data points in the data set, one of the clusters is denser than another. In this case, our approach E_CFSFDP expenses more time to be more robust respect to parameter $d_c$. Fig.7 presents a lot of clusters found by CFSFDP with different $d_c$ and selections of cluster centers. For CFSFDP's performance is depended on the selection of density peaks, we not only present the clustering result, but also show the decision graph and the selections of cluster centers. As shown, CFSFDP gets perfect result when $d_c = \{40\%, 45\%\}$, however it fails with other values, as $d_c = \{1\%, 2\%, 10\%\}$. By more experiments, we find CFSFDP also divides Jain's data set correctly with $d_c = \{44\%, 46\%, 47\%\}$.

As an extension of CFSFDP, our scheme not only can find genuine clusters with the same $d_c$ whereas CFSFDP succeeds (E_CFSFDP can just specify the number of sub-clusters to be 2), but also works fine whereas CFSFDP fails to perform well. E_CFSFDP finds the 2 natural clusters in above data set with parameters of any combination of $d_c = \{4\%, 5\%, 6\%\}$, $N_{neighbor} = \{5, 10, 15\}$, $\beta = \{1, 2, 3, 4, 5\}$. To present the clustering progress better, we present the initial clusters found by E_CFSFDP in Fig. 8. (2), (3), (4).

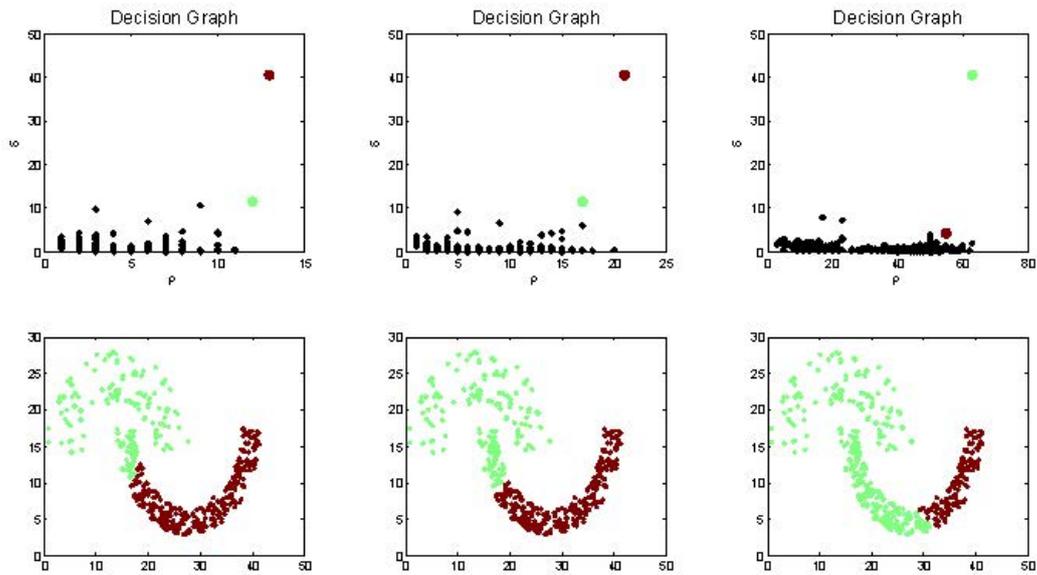

(1) $d_c = 1\%$, 2 clusters. (2) $d_c = 2\%$, 2 clusters. (3) $d_c = 10\%$, 2 clusters.

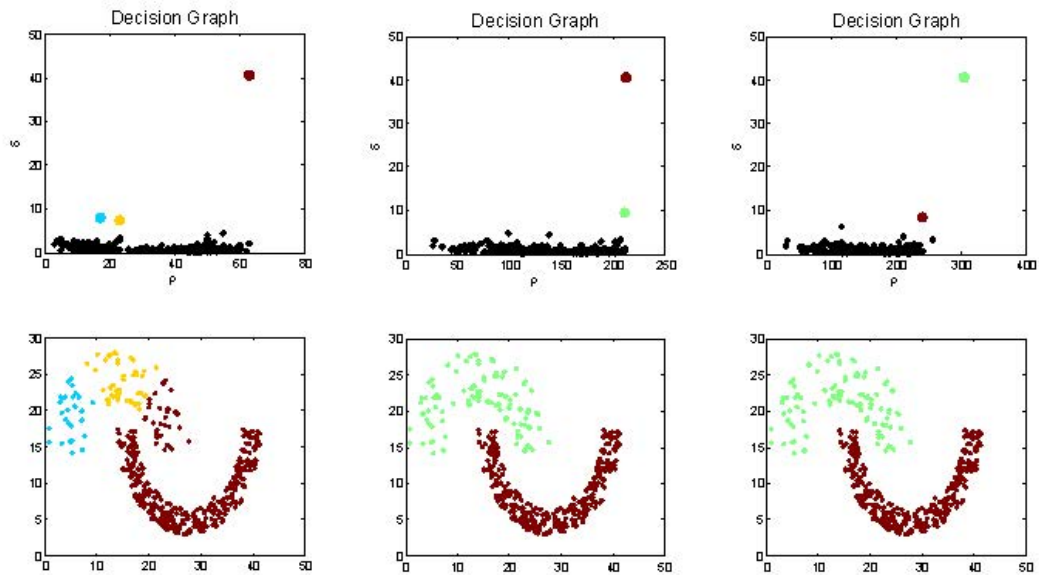

(4) $d_c = 10\%$, 3 clusters. (5) $d_c = 40\%$, 2 clusters. (6) $d_c = 45\%$, 2 clusters.

Figure 7. Clusters found in Jain's data set by CFSFDP with different dc. Different colors correspond to different clusters, there is no any relationship of colors in different figures.

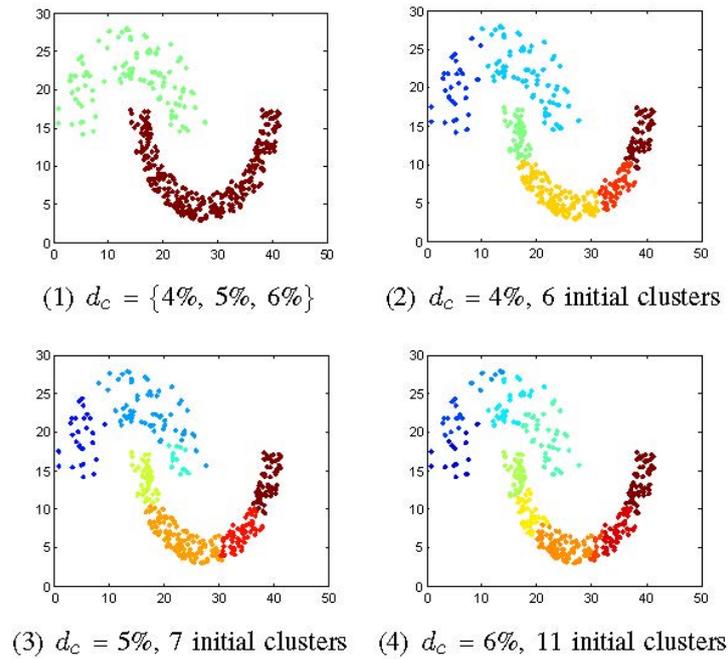

(1) $d_c = \{4\%,\ 5\%,\ 6\%\}$     (2) $d_c = 4\%$, 6 initial clusters

(3) $d_c = 5\%$, 7 initial clusters    (4) $d_c = 6\%$, 11 initial clusters

Figure 8. Clusters found by E CFSFDP in Jain's data set. (1) shows the final results of E CFSFDP. (2), (3), (4) show the initial clusters.

### 2.6.2. Chang's data set

There are 3 natural clusters in the data set, some clustering solutions proposed by CFSFDP have been given in Fig. 4. We also conduct E_CFSFDP on this data set, the result is presented in Fig. 9. In the experiment, we have test CFSFDP and E_CFSFDP on this data set with many parameters, the best result of each approach is given. In this case, E_CFSFDP is the obvious winner.

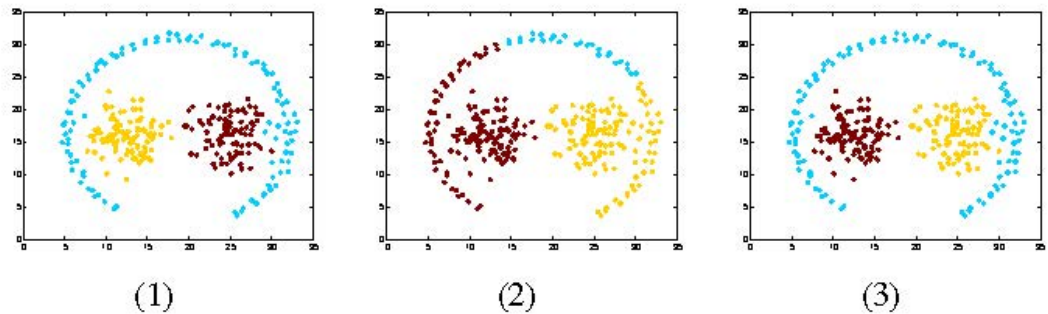

(1)                    (2)                    (3)

Figure 9. Clustering results for Chang's data set. (1) The original data set; (2) CFSFDP clustering result; (3) E CFSFDP clustering result.

Notice that, on this data set, the clustering result of E_CFSFDP is 100% correct, as shown in Fig. 9. (3). That doesn't mean E_CFSFDP is perfect, because in fact, with other values of $d_c$, some points between different clusters will be assigned to wrong cluster.

### 2.6.3. Chameleon 1 and Chameleon 2

There are 8000 points, 6 clusters in Chameleon_1 data set and 10000 points, 9 clusters in Chameleon_2 data set. By experiments, we find that CFSFDP fails to get natural clusters with any value of dc both on the data sets. Fig.10. (1), (2) show the best clustering results we got by

CFSFDP on these data sets. Fig.10. (3), (4) show the best clustering results we got by E_CFSFDP on the two data sets. The two data sets are much denser than the Jain's data set and also Chang's data set, as a result, we use larger values of $N_{neighbor}$ , {30, 35, 40}, smaller values of $d_c$, {1%, 2%, 3%} on these data sets than those of Jain's and Chang's data sets. On these cases, it is clear that E_CFSFDP performs much better than what CFSFDP does.

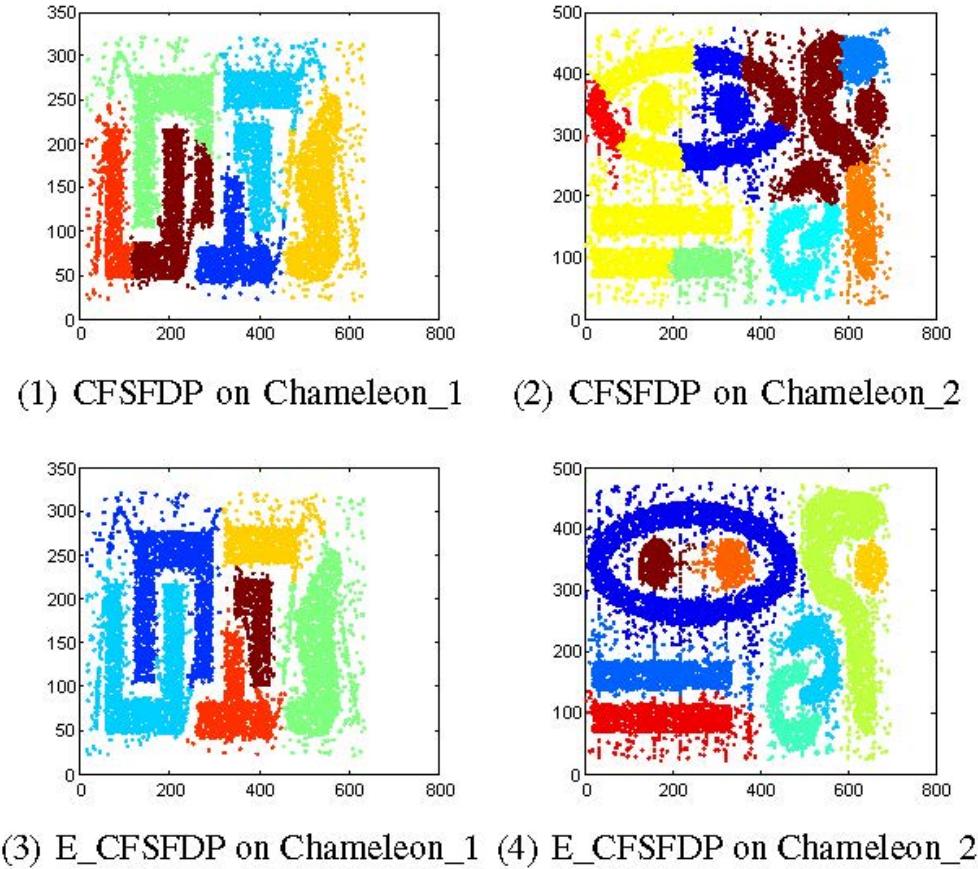

(1) CFSFDP on Chameleon_1    (2) CFSFDP on Chameleon_2

(3) E_CFSFDP on Chameleon_1  (4) E_CFSFDP on Chameleon_2

Figure 10. Clustering results of different schemes for the Chameleon data sets.

# 6. CONCLUSION AND FUTURE WORK

In this paper, we extend a new density-based clustering algorithm CFSFDP [17] to break through the barrier that CFSFDP performs well only when there is unique density peak of each cluster in the data set, what we named as "no density peak". Our solution is taking CFSFDP to get initial clusters, then merge the sub clusters to get final clustering result. The merge progress of our scheme is inspired by the merge phase of CHAMELEON [21]. Instead of the k-nearest neighbor graph, we model the similarity of sub clusters by a variant of k-nearest neighbor graph. Except for that, another difference between our approach and CHAMELEON is that E_CFSFDP is based on density, while CHAMELEON is based on graph partition. In order to demonstrate the applicability of our algorithm to solve the case "no density peak", we conduct CFSFDP and E_CFSFDP on several 2D data sets, there is no density peak for clusters of which. Although our method doesn't increase much more run time complexity than the original algorithm, it's true our algorithm spends more time than the original one.

In future studies, we will focus on reducing the run time of E_CFSFDP. What's more, we will apply it to high dimensional data sets. Another interesting direction is running it in parallel.

## ACKNOWLEDGEMENTS

This work was supported by Key Technologies Research and Development Program of China (No.2012BAH17B03), and the Chinese Academy of Science (No. 2014HSSA09).

**Authors**

**Wen-Kai Zhang** now studies for a M.S degree of School of Computer Science and Technology at University of Science and Technology of China. He received the B.E degree at School of Life Science of USTC. His research mainly concentrated on data mining, anomaly detection.

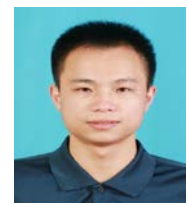

**Jing Li** is a Professor of School of Computer Science and Technology at University of Science and Technology of China. He received the Ph.D degree at USTC in 1993. His research interests include Distributed Systems, Cloud Computing, Big Data progressing and Mobil Computing.

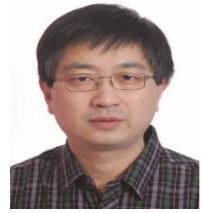